# Realization of locally-round beam in an ultimate storage ring using solenoids

XU Gang (徐刚), JIAO Yi (焦毅), TIAN Saike (田赛克)
Institute of high energy physics, CAS, Beijing 100049, China

## Abstract:
"Ultimate" storage rings (USRs), with electron emittance smaller than 100 pm.rad and on the scale of the diffraction limit for hard X-rays in both transverse planes, have the potential to deliver photons with much higher brightness and higher transverse coherence than that projected for the rings currently operational or under construction. Worldwide efforts have been made to design and to build light sources based on USRs. How to obtain a round beam, i.e. beam with equivalent transverse emittances, is an important topic in USR studies. In this paper, we show that a locally-round beam can be achieved by using a pair of solenoid and anti-solenoid with a circularly polarized undulator located in between. Theoretical analysis and application of this novel method, particularly to one of the Beijing Advanced Photon Source storage ring design having natural emittance of 75 pm.rad, are presented.

## 1.Introduction

"Ultimate" storage rings (USRs) [1], with equal transverse emittances at diffraction limit for X-rays of interest for user community and with much higher performance than existing rings, have been extensively studied in the past few years [2-7].

In a USR, a round beam with equal transverse emittance, i.e, $\varepsilon_x = \varepsilon_y$, is often required. However, associated with the intrinsic horizontally-bending layout of (most of) the storage rings, the vertical dispersion is generally very small or ideally zero. The equilibrium emittance due to quantum excitation and radiation damping usually has much larger portion in the horizontal plane than that in the vertical plane. In this paper, presuming there exists coupling between the horizontal and the vertical plane, we denote the transverse emittances as

$$\varepsilon_x = \frac{\varepsilon_0}{1+\kappa}, \quad \varepsilon_y = \frac{\kappa\varepsilon_0}{1+\kappa}, \tag{1}$$

where $\kappa$ is the coupling factor, which is much smaller than 1 for most of the existing storage rings; $\varepsilon_0 = \varepsilon_x + \varepsilon_y$ is the natural emittance, which is a constant value once the ring optics is fixed.

From Eq. (1), to achieve a round beam, $\kappa = 1$ is required, implying very strong coupling. In this case, $\varepsilon_x = \varepsilon_y = \varepsilon_0/2$, which means the horizontal emittance is reduced by a factor of 2 compared with the general weak coupling case. There are conventionally two ways to produce globally round beam (see Ref. [8] and references therein). One method is to move the tune close to coupling resonance, and the other is to install a Möbius insert. However, the new lattice will be different from that with weak coupling, which inevitably requires re-analysis of the beam dynamics in such a ring. Instead, in this paper we introduce a novel method of producing locally-round beam by using a pair of solenoid and anti-solenoid, with a circularly polarized undulator (called helical undulator) in between. The solenoid (S) introduces strong coupling, and hence is able to produce a round beam with appropriate setting of the solenoid parameters. After the round beam passes through the insertion device (ID) to radiate photons with high brightness and high transverse coherence, the coupling can be perfectly cancelled by using an anti-solenoid (AS) with the same length but opposite field strength. With this S-ID-AS section, the strong coupling region and weak coupling region can be well separated, resulting in a "clean" physics. We notice that there exists a proposal of producing a locally-round beam potentially with a larger emittance-reduction factor by using a solenoid and two triplets of skew quadrupoles [9]. However, this method requires placing insertion device inside the solenoid, and needs a long straight section to install two triplets of skew quadrupoles for decoupling and optics matching. On the other hand, due to the adoption of separate-function elements and less strict requirement for space, our proposed method appears practical and feasible to be fabricated and commissioned in a USR. In addition, we demonstrate that the S-ID-AS layout introduces tolerable perturbations to the global optics, which helps preserve the optimized beam dynamics in the weak coupling case.

The paper is arranged as follows. In Sec. II, we present the theoretical analysis of the beam transportation in an S-ID-AS section, and derive the condition of producing a round beam. In Sec. III, taking the planned light source, Beijing Advanced Proton Source (BAPS) as an example, we show a detailed design to produce a locally-round beam by installing an S-ID-AS section in a 6-m straight section, and perform numerical simulations with ELEGANT code [10]. The agreement between the simulation and theoretical analysis is excellent. Conclusions are given in Sec. IV.

## 2.Beam transport through a solenoid-ID-anti-solenoid section

In this section, we will study the beam transportation through an S-ID-AS section. As mentioned, in an electron storage ring with weak coupling, the beam is flat, i.e., $\varepsilon_x \approx \varepsilon_0$ and $\kappa = \varepsilon_y/\varepsilon_x << 1$. For simplicity, we presume that the particle coordinate is $X_0 = (x_0, x_0', y_0=y_0'=0)$ at the entrance of the solenoid, with $x$ and $x'$ in a general form

$$\begin{aligned} x_0 &= U\cos(\varphi) + V\sin(\varphi), \\ x_0' &= M\cos(\varphi) + N\sin(\varphi). \end{aligned} \tag{2}$$

We set $U$, $V$, $M$, and $N$ in Eq. (2) as arbitrarily real numbers, so as to represent any solution of the transverse betatron motion [11],

$$\begin{aligned} x(s) &= A_x\sqrt{\beta_x(s)}\cos[\psi_x(s)+\psi_{x0}], \\ x'(s) &= -A_x\{\alpha_x(s)\cos[\psi_x(s)+\psi_{x0}]+\sin[\psi_x(s)+\psi_{x0}]\}/\sqrt{\beta_x(s)}, \end{aligned} \tag{3}$$

where $\alpha_x$(s) and $\beta_x$(s) are the horizontal Courant-Snyder parameters as functions of the longitudinal position $s$, $\Psi_x(s)$ is the phase advance of the betatron motion and $\Psi_{x0}$ is the initial phase advance. With straightforward derivation, one can find the corresponding relation between Eq. (2) and Eq. (3),

$$\begin{aligned} &\psi_{x0} = \varphi, \\ &\psi_x(s_0) = \arctan(-V/U), \\ &A_x^2 = |VM - UN|, \\ &\beta_x(s_0) = (U^2+V^2)/|VM-UN|, \\ &\alpha_x(s_0) = -[U+\beta_x(s_0)N]/V, \end{aligned} \tag{4}$$

where $s_0$ indicates the location of the entrance of the solenoid.

In the phase space of ($x$, $x'$), the area $S$ bounded by the particle coordinates in Eq. (2) with specific $U$, $V$, $M$ and $N$ is independent of $\varphi$, $S = |VM-UN|\pi$, which is related to the horizontal emittance by $\varepsilon_x = S/\pi = |VM-UN|$. By contrast, $\varepsilon_y = 0$, corresponding to the presumed initial condition ($y_0$=$y_0$'=0).

Let us consider a flat beam passing through a solenoid, the particle coordinate at the exit of the solenoid $X_1 = (x_1, x_1', y_1, y_1')$ is calculated by

$$X_1 = M_S X_0, \tag{5}$$

where $M_s$ is the 4-by-4 transfer matrix of the solenoid [12],

$$M_S = \begin{bmatrix} M_1^S & M_2^S \\ -M_2^S & M_1^S \end{bmatrix}, M_1^S = \begin{bmatrix} C^2 & SC/k \\ -kSC & C^2 \end{bmatrix}, M_2^S = \begin{bmatrix} SC & S^2/k \\ -kS^2 & SC \end{bmatrix}, \tag{6}$$

where $k = B_z/(2B\rho)$, $C = \cos(kL)$, $S = \sin(kL)$ with $B_z$ being the solenoid field, $B\rho$ being the magnetic rigidity of the central reference trajectory, and $L$ being the effective solenoid length.

From Eqs. (2), (5) and (6), we obtain

$$\begin{aligned} x_1 &= (C^2U + SCM/k)\cos(\varphi) + (C^2V + SCN/k)\sin(\varphi), \\ x_1' &= (-kSCU + C^2M)\cos(\varphi) + (-kSCV + C^2N)\sin(\varphi), \\ y_1 &= -(SCU + S^2M/k)\cos(\varphi) - (SCV + S^2N/k)\sin(\varphi), \\ y_1' &= (kS^2U - SCM)\cos(\varphi) + (kS^2V - SCN)\sin(\varphi). \end{aligned} \tag{7}$$

Taking analogy to the derivation used in the calculation of $\varepsilon_x$ from Eq. (2), we obtain the horizontal and vertical emittances related to Eq. (7),

$$\begin{aligned} \varepsilon_x &= C^2 |VM - UN|, \\ \varepsilon_y &= S^2 |VM - UN|. \end{aligned} \tag{8}$$

The sum of the emittances keeps the same, $\varepsilon_0 = \varepsilon_x + \varepsilon_y = |VM-UN|$. But now the coupling factor $\kappa = \varepsilon_y/\varepsilon_x = T^2$ with $T = S/C = \tan(kL)$, which varies with the value of $kL$. A round beam with $\varepsilon_y = \varepsilon_x = \varepsilon_0/2$ can be obtained by setting $kL = \pi/4$. In this way, we can reduce the horizontal emittance by a factor of 2, and moreover, make the electron emittance of a USR on the level of the diffraction limit for the X-rays of interest in both planes.

If without correction, the strong coupling induced by the solenoid with $kL = \pi/4$ will bring in large perturbations to the global optics and cause difficulty in optimizing the beam dynamics. We will show that using an ID with the same transfer matrixes in the horizontal and vertical planes, such as a helical undulator, that is followed by an AS with the same $L$ but $-k$ (opposite sign), one can perfectly remove the strong coupling.

We represent the ID and AS with 4-by-4 transfer matrixes of

$$M_{ID} = \begin{bmatrix} M_1^{ID} & 0 \\ 0 & M_1^{ID} \end{bmatrix}, M_1^{ID} = \begin{bmatrix} m_{11} & m_{12} \\ m_{21} & m_{22} \end{bmatrix}, \text{ with } m_{11}m_{22} - m_{12}m_{21} \equiv 1. \tag{9}$$

$$M_{AS} = M_S(k \to -k). \tag{10}$$

Then we can obtain the transfer matrix of the S-ID-AS section,

$$M_{S-ID-AS} = M_{AS}M_{ID}M_S = \begin{bmatrix} R & 0 \\ 0 & R \end{bmatrix}, R = \begin{bmatrix} r_{11} & r_{12} \\ r_{21} & r_{22} \end{bmatrix}, \tag{11}$$

with

$$
\begin{aligned}
r_{11} &= [km_{11}C^2 + (-k^2 m_{12} + m_{21})CS - km_{22}S^2]/k, \\
r_{12} &= [k^2 m_{12}C^2 + k(m_{11} + m_{22})CS + m_{21}S^2]/k^2, \\
r_{21} &= m_{21}C^2 - k(m_{11} + m_{22})CS + k^2 m_{12}S^2, \\
r_{22} &= [km_{22}C^2 + (-k^2 m_{12} + m_{21})CS - km_{11}S^2]/k.
\end{aligned}
\tag{12}
$$

Note that $M_{S\text{-}ID\text{-}AS}$ has only nonzero diagonal 2-by-2 matrixes. Thus, after passing through an S-ID-AS section, the particle coordinate at the exit of the anti-solenoid, $X_2 = M_{S\text{-}ID\text{-}AS}X_0 = (x_2, x_2', y_2 = y_2' = 0)$, with

$$
\begin{aligned}
x_2 &= (r_{11}U + r_{12}M)\cos(\varphi) + (r_{11}V + r_{12}N)\sin(\varphi), \\
x_2' &= (r_{21}U + r_{22}M)\cos(\varphi) + (r_{21}V + r_{22}N)\sin(\varphi).
\end{aligned}
\tag{13}
$$

The horizontal emittance is, again, $\varepsilon_x = \varepsilon_0 = |VM-UN|$, implying perfect cancellation of the strong coupling induced by the solenoid.

We find that the transfer matrix $M_{S\text{-}ID\text{-}AS}$ can be decomposed into a combination of drift and thin-lens "quadupole", which provides focusing in both planes,

$$
M_{S-ID-AS} = M_D M_Q M_Q M_D, \tag{14}
$$

with

$$
M_D = \begin{bmatrix} M^D & 0 \\ 0 & M^D \end{bmatrix}, M^D = \begin{bmatrix} 1 & L_d \\ 0 & 1 \end{bmatrix}, M_Q = \begin{bmatrix} M^Q & 0 \\ 0 & M^Q \end{bmatrix}, M^Q = \begin{bmatrix} 1 & 0 \\ -1/f & 1 \end{bmatrix}.
$$

where $f = -2/r_{21}$, and $L_d = (r_{11} - 1)/r_{21}$.

From Eq. (14), the S-ID-AS section will introduce additional focusing and cause tune shift to the ring, which, however, can be corrected by using quardupoles near ID. In next section, we illustrate the concrete application of this novel method in the 5 GeV storage ring-based light source, BAPS.

## 3.Solenoid-ID-anti-solenoid section in BAPS

A storage ring-based light source, Beijing Advanced Photon Source (BAPS) is planned to be built in Beijing to satisfy the increasing requirements of high-brilliance, high-coherence radiation from the user community. One lattice of the BAPS storage ring, with circumference of 1263.4 m and natural emittance of 75 pm, is recently proposed as one candidate of the final design [7]. The ring consists of 16 superperiods, and each superperiod contains one high-beta 10-m straight section for injection and one low-beta 6-m straight section for IDs. Table 1 lists the main parameters of the ring and Fig. 1 shows the optical functions in one superperiod. With a combination of several progressive technologies, such as quasi-3rd-order achromat [6], theoretical analysis based on Lie Algebra and Hamiltonian dynamics [7], multi-objective genetic algorithm [13], we obtain large enough dynamic aperture for pulsed sextupole injection [14]. Figure 2 shows the optimized dynamic aperture and frequency map in the case of weak coupling, the dynamic aperture is larger than 8 mm in the horizontal plane.

In one low-beta 6-m straight section, we adopt solenoid and anti-solenoid with a length of 1.75 m and a field of 14.97 Tesla, and place a 2.5-m helical undulator with undulator period of 2.5 cm and a peak field of 0.5 Tesla in between. The used solenoid field is large but feasible with today's technologies [15]. Considering only the linear effects, the diagonal 2-by-2 matrix $R$ of the matrix $M_{S\text{-}ID\text{-}AS}$ [see Eq. (11)] is

$$R = \begin{bmatrix} 0.998596 & 2.49883 \\ -0.0011229 & 0.998596 \end{bmatrix} \tag{15}$$

From Eqs. (14) and (15), this is equivalent to replacing the 6-m straight section with two drifts of 7.93 m and two thin-lens quadupoles with focusing length of 10.15 m. Using this equivalent model, we can easily check the linear effect of the S-ID-AS section to the ring optics. We find that if without correction, the linear optics becomes unstable. To recover the tune, we vary the quadrupole strengths (three variables) of the two nearby triplets for optics matching. Since the variables are not enough, only the horizontal tune is recovered, the vertical tune after correction increases by 0.12. Numerical tracking (see Fig. 3) shows that the horizontal dynamic aperture is reduced to about 4 mm, which is still satisfactory for the pulsed sextupole injection.

In practice, one can completely correct the tune shift by using more variables. In that case, the phase advance between the chromatic and harmonic sextupoles in the ring may be changed, and re-optimization of the nonlinear beam dynamics will be required. This topic is not so much relevant to the main issue of this paper, so it will not be addressed here.

Table 1: Main parameters for the BAPS storage ring with natural emittance of 75 pm.rad

| **Parameter** | **Flux** | **Unit** |
|---|---|---|
| Energy | 5 | GeV |
| Circumference | 1263.4 | m |
| Horizontal damping partition number $J_x$ | 1.40 | |
| Natural emittance | 75 | pm |
| Working point (H/V) | 98.4/34.3 | |
| Natural chromaticities (H/V) | -189/-113 | |
| Number of superperiods | 16 | |
| Number of high-beat 10-m straight sections | 16 | |
| Beta functions in high-beta straight section (H/V) | 41/4.7 | m |
| Number of low-beta 6-m straight sections | 16 | |
| Beta functions in low-beta straight section (H/V) | 4.5/1.7 | m |
| Damping times (*x/y/z*) | 20/28/17.4 | ms |
| Energy spread | $8\times10^{-4}$ | |
| Momentum compaction | $3.86\times10^{-5}$ | |

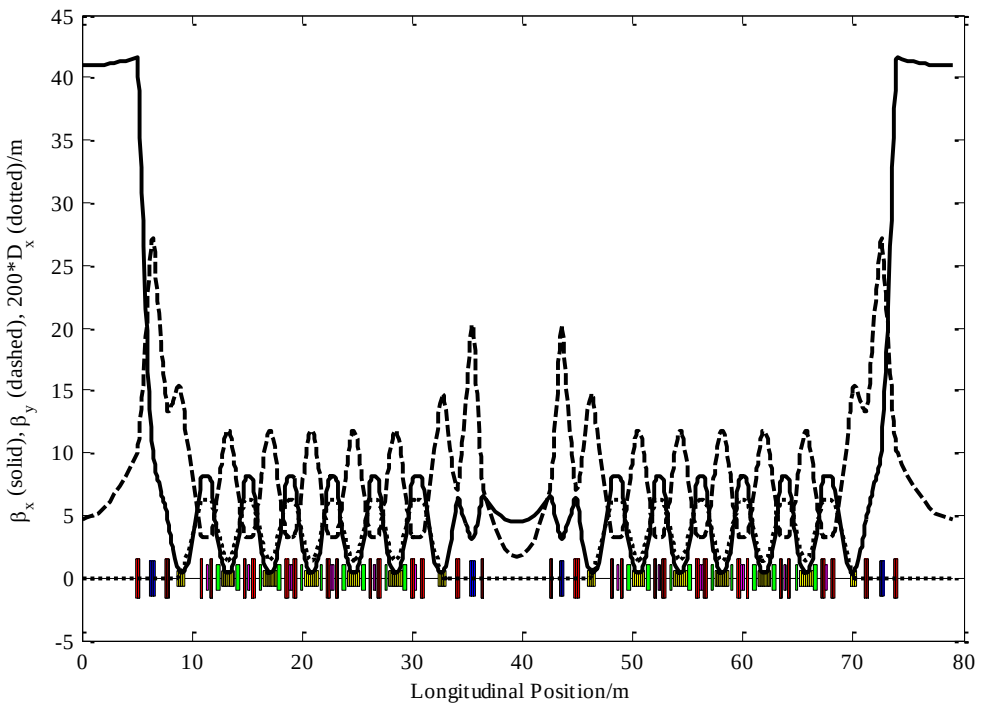

Fig. 1. Optical functions in a superperiod of the designed BAPS storage ring.

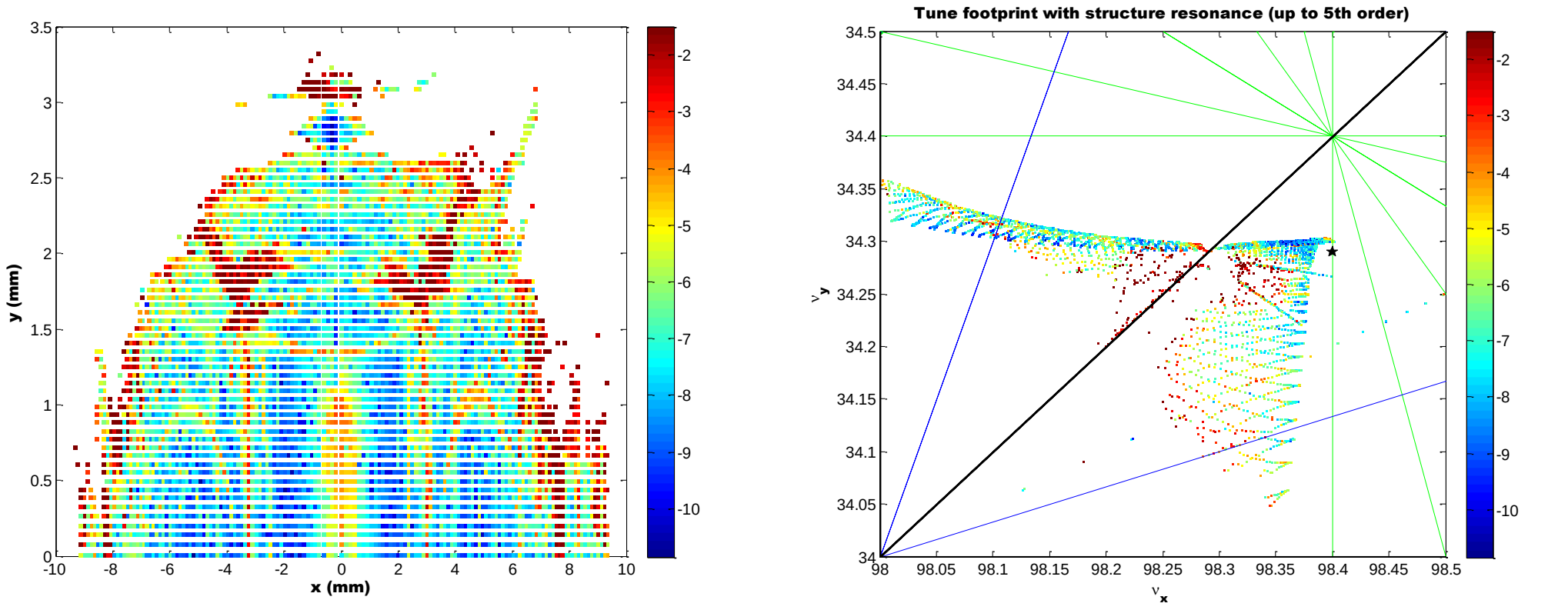

Fig. 2. (color online) Dynamic aperture and frequency map for BAPS storage ring with weak coupling.

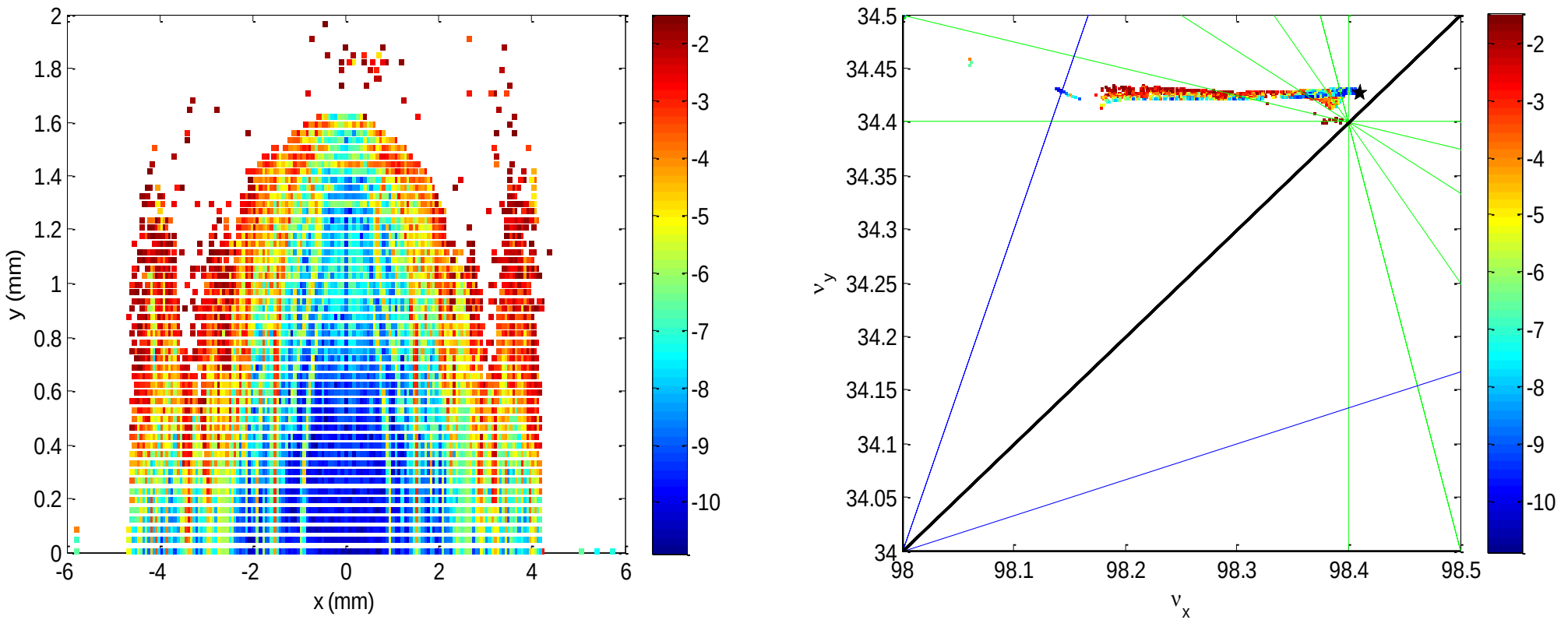

Fig. 3. (color online) Dynamic aperture and frequency map for BAPS storage ring with S-ID-AS section and correction of the tune shift.

To verify the theoretical analysis, we use ELEGANT code to perform multi-particle tracking for the designed BAPS storage ring with the S-ID-AS section and corrected triplet quadrupoles. In the simulation, the solenoid is implemented as a matrix, up to the 2nd order; the helical undulator is presumed to have an ideal sinusoidal field, and is represented with a number of canonical kicks; a 498.8 MHz, 6 MV RF cavity and synchrotron radiation effects are also considered. We set the initial electron beam dimensions to somewhat arbitrarily chosen values, $\varepsilon_x$ = 7 nm.rad, $\varepsilon_y$ = 0.18 nm.rad, $\sigma_z$ = 4 mm and $\sigma_\delta = 7\times10^{-4}$, and then track the particles for 35 thousand turns, during which no particle loss is observed. The evolution of the transverse emittances and the phase space distributions at the two ends and in the middle of the S-ID-AS section are shown in Figs. 4 and 5, respectively. After three damping times (about 20 thousand turns), the transverse emittances reach equilibrium values; in the middle of the section, the electron beam has similar beam dimensions and the same emittances ($\varepsilon_x$ = $\varepsilon_y$ ~ 37.5 pm.rad) in both transverse planes, in contrast to the other two locations where $\varepsilon_x$ ~ 75 pm.rad and $\varepsilon_y$ << $\varepsilon_x$. The simulation result agrees with the prediction of the theoretical analysis very well.

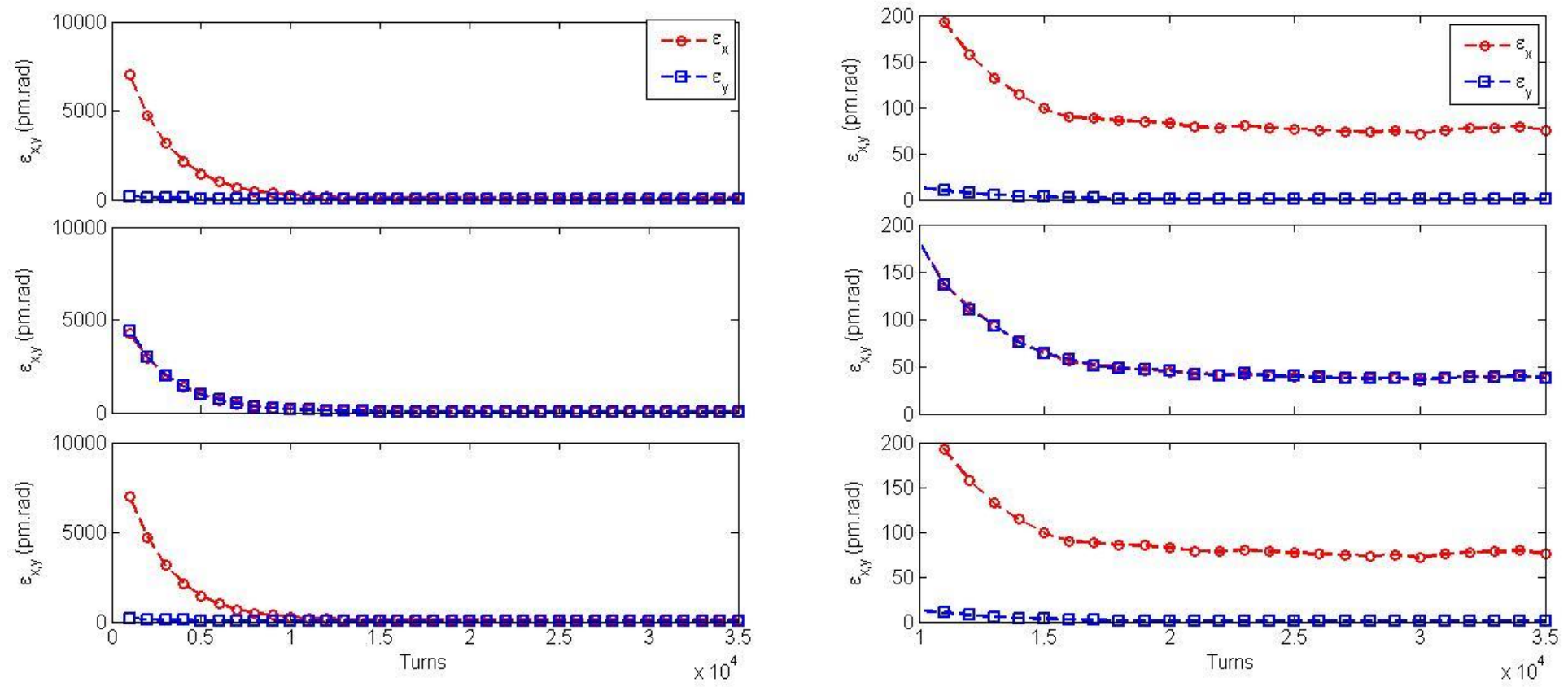

Fig. 4. Evolution of the horizontal and vertical emittances of the electron beam at the entrance of the solenoid (upper), in the middle of the undulator (middle), and at the exit of the anti-solenoid (bottom). The right plot highlights the emittances approaching equilibrium values.

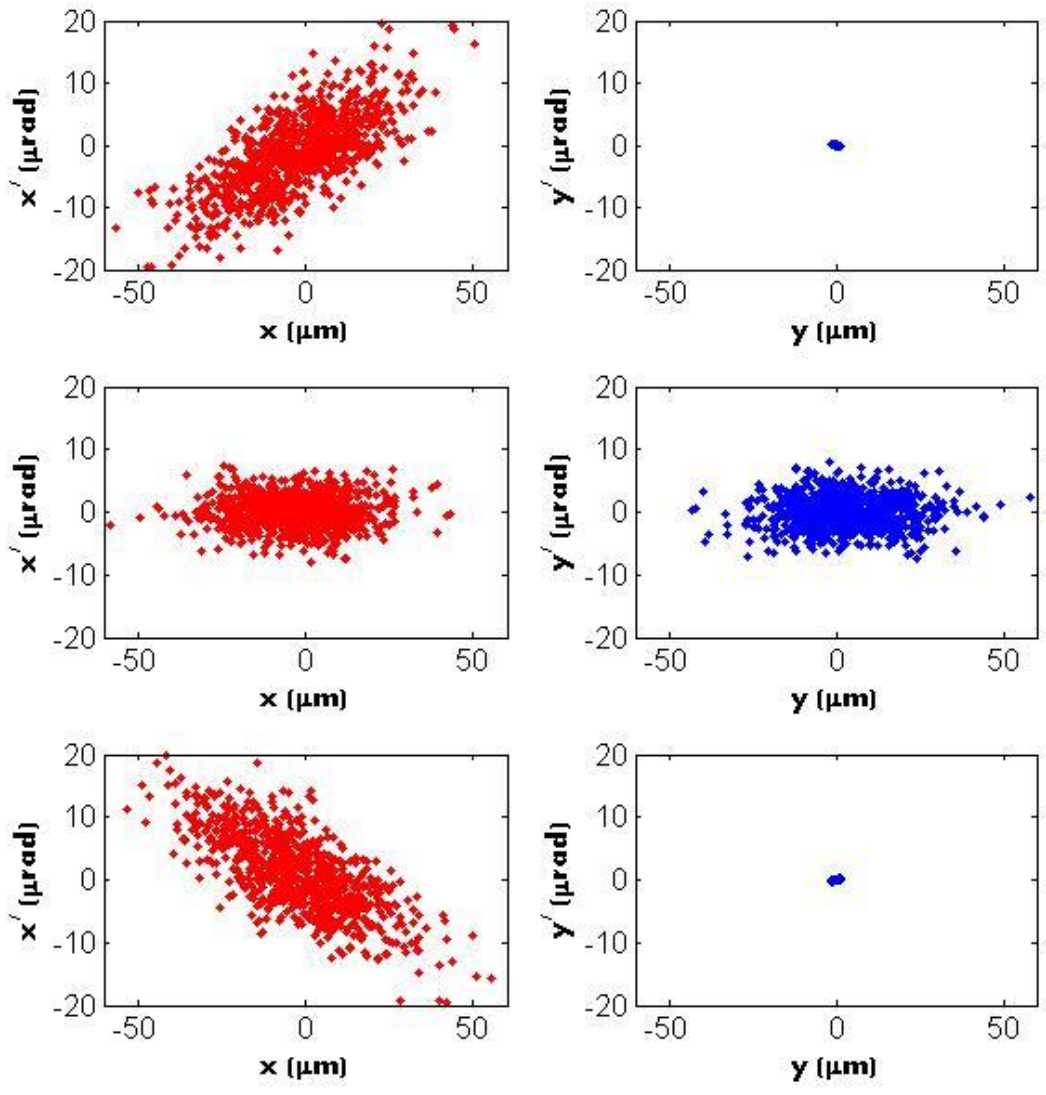

Fig. 5. The transverse phase space distributions at the entrance of the solenoid (upper), in the middle of the undulator (middle), and the exit of the anti-solenoid (bottom).

## 4.Conclusions

In this paper, we propose and demonstrate that one can use a solenoid (S) and an anti-solenoid (AS) to produce a locally round beam, i.e., $\varepsilon_x = \varepsilon_y = \varepsilon_0/2$, in the insertion device (ID) locating between the solenoids, and therefore improve the quality of the radiated photon beam. Installing such a section in an ultimate storage ring can induce a net tune shift, which, however, can be compensated by using the nearby quadrupole magnets. Above all, this study shows a novel method to produce round beam in a storage ring, as required by an ultimate storage ring.

**Acknowledgement**

We appreciate Prof. A. Chao of SLAC and Prof. WANG Jiuqing of IHEP for helpful discussions. This work was supported by the special fund of the Chinese Academy of Sciences under contract No. H9293110TA.